\documentclass{article}  
\usepackage{breckenr2005}
\usepackage{graphicx}
\frompage{000} \topage{000}                                              %|
\def\rightmark{Scaling properties of azimuthal anisotropy of mesons and baryons}
\def\leftmark{A. Taranenko for the PHENIX Collaboration}
\def\pT{\mbox{$p_T\!$ }}
\def\v2{\mbox{$v_2$}}
\def\sqrtsNN{\mbox{$\sqrt{s_{NN}}$}} 
\title{Scaling properties of azimuthal anisotropy of mesons and baryons at RHIC} 
\authors{
{Arkadij Taranenko$^1$ for the PHENIX Collaboration %
}\\[2.812mm]
{\normalsize
\hspace*{-8pt}$^1$ Department of Chemistry, SUNY Stony Brook, \\ 
11790  Stony Brook, USA\\[0.2ex] 
}}
 
\abstract{Detailed measurements of the azimuthal anisotropy (\v2) for identified charged particles
are reported as a function of transverse momentum ($p_T$) and centrality for Au+Au
collisions at \sqrtsNN~=200~GeV. The measurements indicate clear evidence for eccentricity and 
particle flavor scaling over a broad range of centralities and transverse rapidity $y_T$, indicating 
a hydrodynamical origin of the fine structure of azimuthal anisotropy at RHIC.
The observed scaling supports the picture of a suddenly hadronizing (recombining) fluid of quarks. 
An apparent breaking of flavor scaling at relatively large values of $y_T$ points to 
an important change in the mechanism for particle emission. 
}
\PACS{specifications see, e.g.\ {\tt http://www.aip.org/pacs/}}
 
\begin{document}
 
\maketitle
\setcounter{page}{1}

\section{Introduction}\label{intro}

During the early stages of an ultra-relativistic heavy-ion collision, an extremely 
high energy density system, possibly consisting of a new phase of nuclear matter, 
is expected to be formed~\cite{Shuryak:2004cy,Gyulassy_qgp,Muller:2004kk,qgp1,qgp2}.
The dynamical evolution of this matter is predicted to reflect its 
properties~\cite{Shuryak:2004cy,Gyulassy_qgp,Muller:2004kk}. Consequently,
much effort is currently centered on the study of reaction dynamics at the Relativistic 
Heavy Ion Collider (RHIC). Azimuthal correlation measurements constitute an important 
probe for reaction dynamics. They serve as a ``barometric sensor" for pressure 
gradients developed in the collision and hence yield insight into crucial issues 
of thermalization and the equation of state (EOS)~\cite{Ollitrault:1992bk,Kolb:2001qz,Hirano:2004rs}.
They provide important constraints for the density of the medium and the 
effective energy loss of partons which traverse it~\cite{Gyulassy:2000gk,Molnar:2001ux}.
They can even provide valuable information on the gluon saturation scale in the 
nucleus~\cite{Kovchegov:2002nf}.
  
  	Azimuthal correlation measurements show significant harmonic strength at mid-rapidity with 
characteristic dependencies on \pT and centrality~\cite{star1,flow04,Ajitanand:2002qd,ppg022}.
The anisotropy of this harmonic pattern is typically characterized by the second order Fourier coefficient, 
\begin{eqnarray}
 v_2 = \left\langle e^{i2(\phi_1-\Phi_{RP})} \right\rangle,
\label{eq1}
\end{eqnarray}
where $\phi_{1}$ represents the azimuthal emission angle of an emitted particle 
and $\Phi_{RP}$ is the azimuth of the reaction plane. The brackets denote statistical 
averaging over particles and events. A large amount of available data for Au+Au collisions at 
\sqrtsNN~=~62.4, 130, and 200~GeV \cite{star1,flow04,ppg022,star2,ppg047} indicate that 
the magnitude and trends of \v2 (for $p_T \mathbin{\lower.3ex\hbox{$\buildrel<\over
{\smash{\scriptstyle\sim}\vphantom{_x}}$}} 2.0$ GeV/$c$) are under-predicted by hadronic cascade 
models supplemented with string dynamics~\cite{Bleicher:2000sx}, 
but are well reproduced by models which incorporate hydrodynamic flow~\cite{Shuryak:2004cy,Kolb:2001qz}.
This has been interpreted as evidence for the production of a thermalized state of 
partonic matter~\cite{Shuryak:2004cy,Gyulassy_qgp,Muller:2004kk}. If this is indeed the case, then 
the fine structure of azimuthal anisotropy (ie. its detailed dependence on centrality, 
transverse momentum, particle type, higher harmonics, etc) should reflect the 
scaling ``laws" predicted by ideal hydrodynamics.

	In this work we use detailed differential \v2 measurements to test for such 
scaling ``laws" and the onset of competing mechanisms.

\section{Hydrodynamic scaling}

	 An important scaling prediction of hydrodynamic theory is exemplified by the exact analytic hydro 
solutions~\cite{Csorgo:2001xm} exploited in the Buda-Lund model~\cite{buda-lund}.
For harmonic flow, the model gives:
\begin{equation}
v_{2n} = \frac{I_n(w)}{I_0(w)}, n = 1,2,.., \  w = \frac{p_t^2}{4\overline{m}_t} 
\left(\frac{1}{T_{y}} - \frac{1}{T_{x}}\right),
\label{eq2}
\end{equation}
where $I_{0,n}$ are modified Bessel-functions, $\overline{m}_t$ is an                      
average of the rapidity dependent transverse mass (at mid-rapidity,       
$\overline{m_T} = m_T$), and $T_{x}$ and  $T_{y}$ are 
direction ($x$ and $y$) dependent slope parameters:
\begin{eqnarray}
   T_{x}&=&T_0+\overline{m}_t \, \dot X_f^2
       \frac{T_0}{T_0 +\overline{m}_t a},\\
   T_{y}&=&T_0+\overline{m}_t \, \dot Y_f^2
     \frac{T_0}{T_0 +\overline{m}_t a}.
\label{eq3}
\end{eqnarray}
Here, $\dot X_f$ and $\dot Y_f$ gives the transverse expansion rate of the 
fireball at freeze-out, and $a = (T_0-T_s)/T_s$ is its transverse 
temperature inhomogeneity, characterized by the temperature at its    
center $T_0$, and at its surface $T_s$. The important prediction of 
Eq.~\ref{eq2} is that the relatively complicated fine structure of azimuthal 
anisotropy can be scaled to a single function. An illustration of this fact 
can be made for particle flavor scaling via substitution of the transverse 
rapidity \cite{buda-lund2}, $y_T=sinh^{-1}(p_T/m)$ in Eq.~\ref{eq2} to give 
\begin{equation}
v_2=k_1 y_T^2\frac{m}{T_0}\left(1+\frac{k_2}{k_1}\frac{T_0}{m}+\frac{k_3}{k_1}\frac{T_0^2}{m}...\right),
\label{eq4}
\end{equation}
where, k1, k2, k3... are largely governed by the expansion rate. Close inspection of  
the leading term in Eq.~\ref{eq4} indicates that $v_2$ for different particle species 
should scale with $y_T^{fs} = k_m\cdot y_T^2 m$. Here $k_m$ is a mass dependent factor 
which is $\simeq 1$ for relatively heavy particles. It is straightforward to show 
that hydrodynamics also predicts that $v_2$ should scale with the spatial 
eccentricity $\varepsilon =( Y^2- X^2)/( Y^2+ X^2)$, of the overlap between the two
colliding gold nuclei and the higher harmonic $v_4 \sim \frac{1}{2}v_2^2 + k_m y_T^4$. 

%In the ensuing sections we present the results of detailed  $v_2(p_T, centrality)$  
%measurements for charged 
%mesons ( $\pi^{\pm}$, K$^{\pm}$ ) and baryons ( p , $\bar{p}$ ) from   Au+Au collisions 
%at \sqrtsNN~=~200 GeV   and test for several scaling rules.

\section{Analysis}\label{techno}  

	The present analysis is based on $\simeq 22$~M minimum bias Au+Au events obtained with the 
PHENIX detector at \sqrtsNN~=~200 GeV during the second running period (2002) at RHIC.   
Charged tracks were detected in the east and west central arms of PHENIX~\cite{phenix,run02}, 
each of which subtends 90$^\circ$ in azimuth $\phi$, and $\pm 0.35$ units of pseudo-rapidity $\eta$. 
Track reconstruction was accomplished at each collision energy via pattern recognition using 
a drift chamber (DC) followed by two layers of multi-wire proportional
chambers with pad readout (PC1 and PC3) located at radii of 2 m, 2.5 m and 5 m 
respectively~\cite{phenix,run02}. The collision vertex $z$ along the beam direction was 
constrained to be within $|z| <$~30~cm.  A confirmation hit within a $2\sigma$ 
matching window was required in PC3 and the electromagnetic calorimeter (EMC PbSc) or the 
time-of-flight detector (TOF), to eliminate most albedo, conversions, and decays. 
Particle momenta were measured with a resolution of $\delta p/p = 0.7\% \oplus 1.0\%\,p$~(GeV/$c$). 

	Event centralities were obtained via a series of cuts in the space of BBC 
versus ZDC analog response; they reflect percentile cuts on the total 
interaction cross section~\cite{Adcox:2003nr}. 
Estimates for the number of participant nucleons N$_{part}$, were 
also made for each of these cuts following the Glauber-based model detailed 
in Ref.~\cite{flow04}.

	In this analysis the combination of the TOF detector and six sectors of the (EMC PbSc) was 
used to identify charged particles. Particle time-of-flight was measured using the TOF ( or EMC ) 
and the collision time defined by beam counters (BBC). A timing resolution of $\simeq$ 120 ps  
and $\simeq$ 370-400 ps was obtained for the TOF and the EMC (PbSc) respectively. This allowed
meson ( $\pi^{\pm}$, K$^{\pm}$ ) and baryon ( p , $\bar{p}$ ) separation up 
to a $p_T \simeq $ 3.5 GeV/c with the TOF detector and up to 2.5 GeV/c with the  (EMC PbSc). 

The differential $v_2(p_T, centrality)$ measurements were obtained via the reaction plane 
technique which correlates the azimuthal angles of charged tracks detected in the central arms
with the azimuth of an estimated event plane $\Phi_2$, determined via hits in the North and South
BBC's located at $\mid{\eta}\mid \sim 3 - 3.9$~\cite{ppg022,rplane}. 
Due  the large rapidity gap ($\sim 3 - 3.9$ units) between the particles used for 
reaction plane determination and the mid-rapidity particles
correlated with this plane, one  expects  that the latter correlations are less 
influenced by non-flow contributions especially for $p_T < 3.0$~GeV/$c$.
The estimated resolution of the combined reaction plane from both BBC's~\cite{ppg022,rplane} 
has an average of 0.33 over centrality with a maximum of about 0.42 
for Au+Au collisions at  \sqrtsNN=~200~ GeV. Thus, the estimated correction factor, 
which is the inverse of the resolution for the combined reaction plane, 
ranges from 2.4  to 5.0.

\subsection{$v_2$ results}\label{details}
\begin{figure}[htb!]
\includegraphics[width=0.98\linewidth]{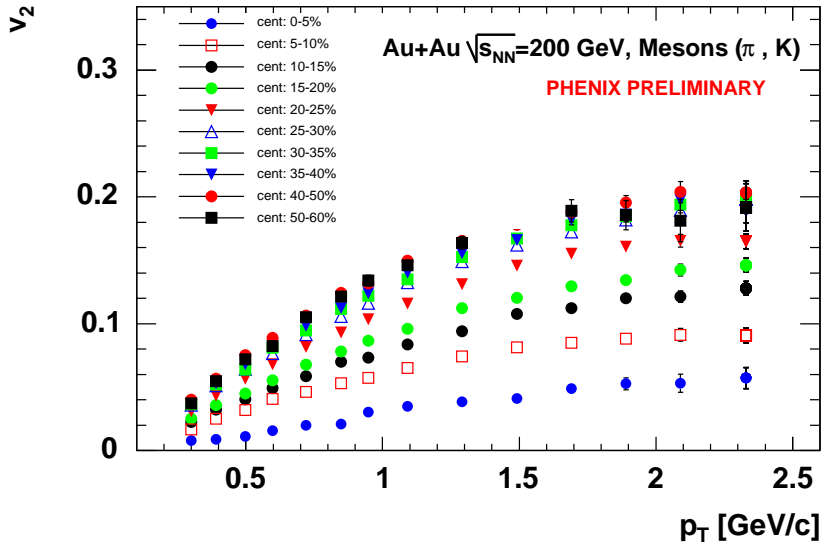}
\vskip -0.7cm
\caption[]{ v$_{2}$  vs  $p_T$  for charged mesons ( $\pi^{\pm}$, K$^{\pm}$ ) for 
different bins in reaction centrality (top to bottom) 0-5\%,5-10\%, ... , 35-40\%, 
40-50\%, 50-60\%. The error bars shown indicate statistical errors only. Systematic 
errors are estimated to be less than 10\%.
\label{v2meson}
}
\end{figure}

\begin{figure}[htb!]
\includegraphics[width=0.98\linewidth]{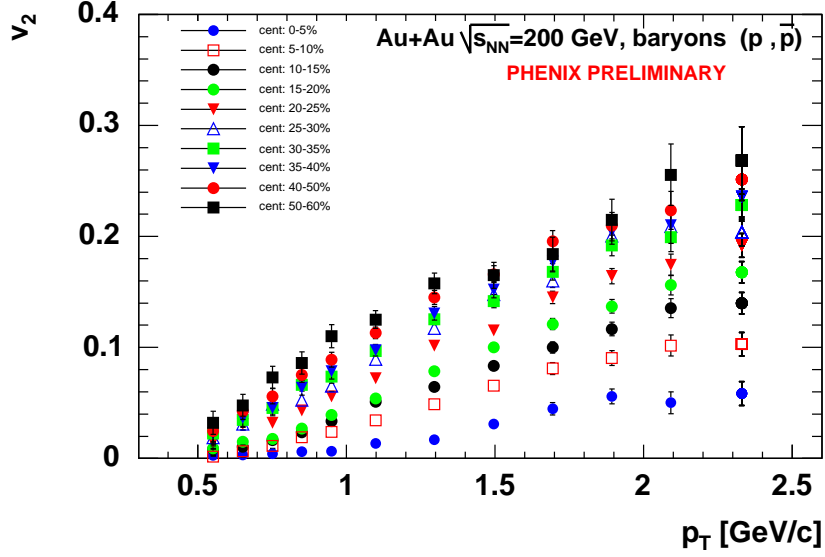}
\vskip -0.7cm
\caption[]{ v$_{2}$  vs  $p_T$  for charged baryons ( p + $\bar{p}$ ) for 
different bins in reaction centrality (top to bottom) 0-5\%,5-10\%, ... , 35-40\%, 
40-50\%, 50-60\%. The error bars shown indicate statistical errors only. Systematic 
errors are estimated to be less than 10\%.
\label{v2baryon}
}
\end{figure}

Figures~\ref{v2meson} and \ref{v2baryon} summarize the centrality and $p_T$ dependence of $v_2$ for
charged mesons( $\pi^{\pm}$, K$^{\pm}$ )  and  baryons ( p + $\bar{p}$ ) detected in the EMC+TOF 
respectively.
 They give an excellent overview of the the evolution of $v_2$ as centrality and $p_T$ 
are varied. The results shown for protons and anti-protons give an especially good view of the 
evolution away from the well know quadratic dependence of $v_2(p_T)$ 
(which is also observed in very central collisions for these data) as the collisions 
become more peripheral. Such a dependence could result from changes in the freeze-out 
temperatures and/or the radial flow velocity as the collisions become more peripheral.

\subsection{Eccentricity scaling}\label{ecc_scaling}

	As indicated earlier, hydrodynamics \cite{kolb}  predicts eccentricity scaling of the azimuthal 
anisotropy. To test for this scaling, one can divide the $v_2$ values obtained at a 
given centrality by the eccentricity $\epsilon$ obtained from a Glauber-based calculation 
obtained for the same centrality \cite{flow04}. Alternatively, one
can simply divide the differential anisotropy $v_2(N_{part}, p_T)$ obtained at a 
given centrality by the $p_T$ integrated value for the same centrality 
selection  $v_2(N_{part})$. The underlying idea here is that the integral flow 
is monotonic and linearly related to the eccentricity over a broad range of 
centralities \cite{heiselberg}. Another advantage of this approach 
is that the ratio  $v_2(N_{part}, p_T)$/$v_2(N_{part})$ gives 
a scale invariant variable which automatically reduces the systematic errors associated 
with an eccentricity evaluation, and the reaction plane resolution. 

\begin{figure}[htb!]
\includegraphics[width=0.98\linewidth]{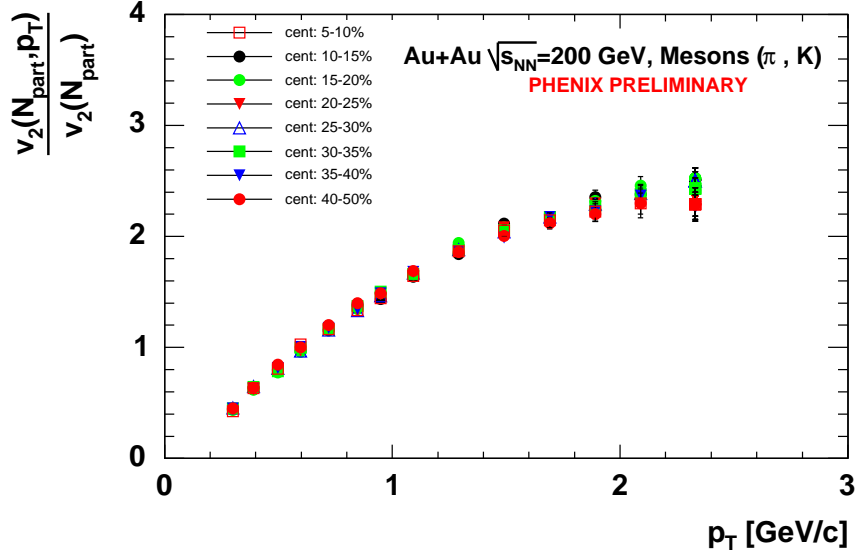}
\vskip -0.65cm
\caption[]{ v$_{2}$  vs  $p_T$  for charged mesons ( for different bins in 
reaction centrality ) scaled by the value of integral flow 
for each centrality..
\label{v2mesonscale}
}
\end{figure}

\begin{figure}
\begin{minipage}[htb!]{0.53\linewidth}
%\centering
%\begin{flushleft}
\hspace*{-.55cm}
\includegraphics[width=1.1\linewidth]{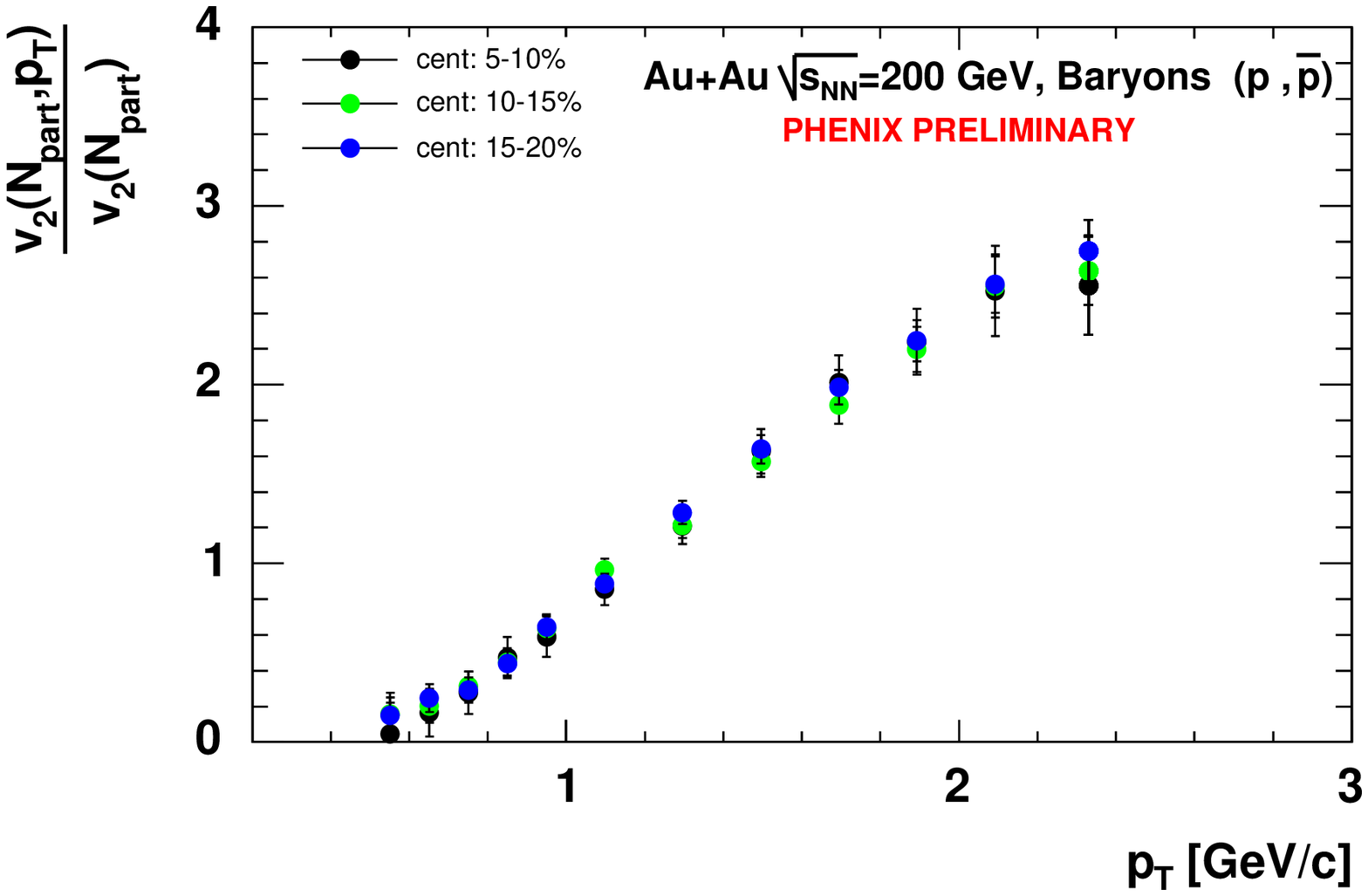}
\vskip -1.2cm
%\caption{\small{ xxxx}}
%\label{auau_cen}
%\end{flushleft}
\end{minipage}
%\hskip 0.2cm
\begin{minipage}[htb!]{0.53\linewidth}
%\centering
%\begin{flushright}
\hspace*{-.55cm}
\includegraphics[width=1.1\linewidth]{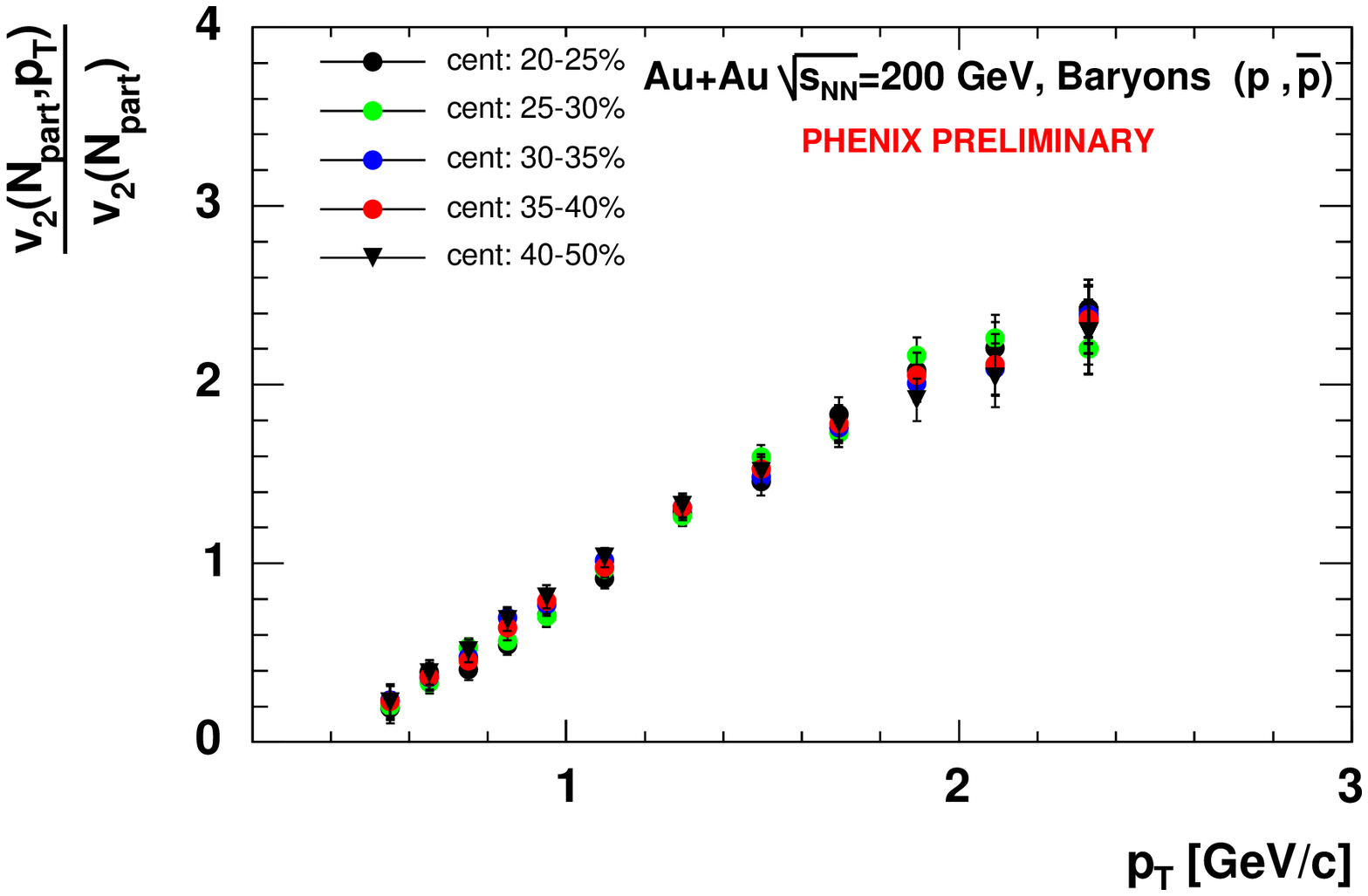}
\vskip -1.2cm
%\caption{\small{xxxx}}
%\label{auau_kt}
%\end{flushright}
\end{minipage}
\vskip -0.5cm
\caption{v$_{2}$  vs  $p_t$  for charged baryons ( for different bins in 
reaction centrality ) scaled by the value of integral flow 
for each centrality. Left panel corresponds to centrality range 0-20\% 
and right panel to centrality range 20-50\%  
\label{v2baryonscale}
}
\end{figure}

Fig.~\ref{v2mesonscale} shows the v$_{2}$ of charged mesons obtained for different 
centralities scaled by the value of the integral flow obtained for each of these 
centralities. The figure shows essentially perfect scaling for mesons as would be 
expected for a process driven largely by the eccentricity of the overlap
region of the two colliding nuclei. 
The results of a similar scaling test for baryons are shown in Fig.~\ref{v2baryonscale}.
Despite a continuous evolution in the shape of v$_{2}$  vs  $p_T$ with centrality, they 
indicate a relatively good scaling for the two centrality ranges: 0-20\% (~left panel~) 
and 20-50\% (~right panel~).

\subsection{Flavor scaling}\label{flavor_scaling}
\begin{figure}[htb!]
\hspace*{-1.1cm}
\includegraphics[width=1.12\linewidth]{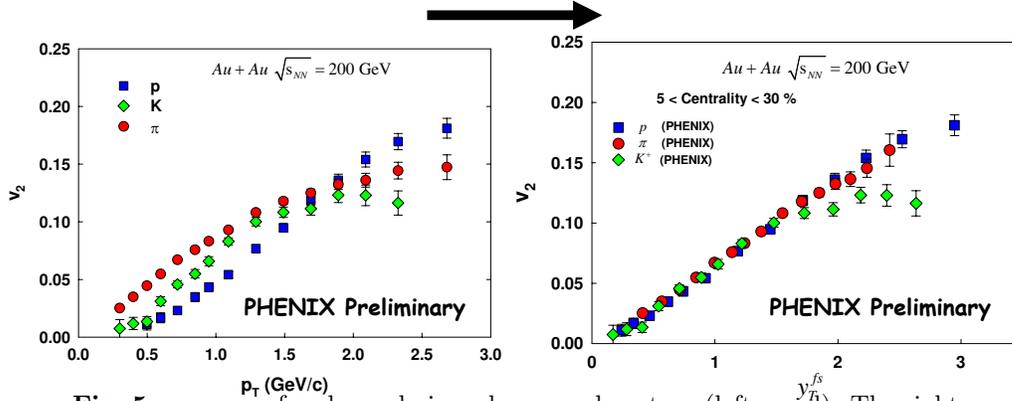}
\vskip -0.8cm
\caption[]{ v$_{2}$  vs  $p_T$  for charged pions,  kaons and protons (left panel). 
The right panel shows the result of the $y_T^{fs}$ scaling predicted by 
hydrodynamics. Results are shown for the centrality selection 5-30\%.
\label{fig5}
} 
\end{figure}

\begin{figure}[htb!]
%\hspace*{0.15cm}
\includegraphics[width=1.0\linewidth]{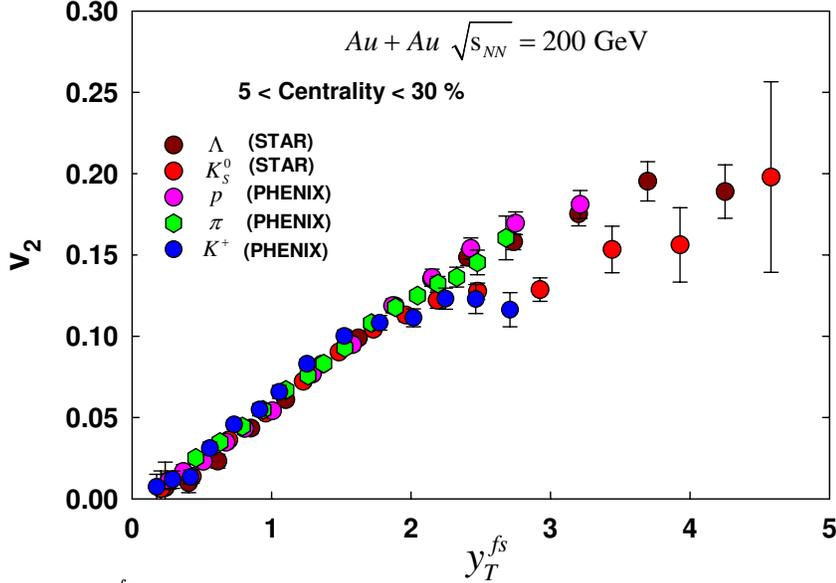}
\vskip -0.8cm
\caption[]{ y$_{T}^{fs}$  scaling of charged pions , kaons and protons (PHENIX) and 
neutral kaons and lambdas (STAR) \cite{star2}. Results are shown for the centrality selection 5-30\%.
\label{fig6}
} 
\end{figure}

	Following Eq.~\ref{eq4}, the scaling variable $y_T^{fs}=k_m\cdot y_T^2 m$, was used to 
investigate flavor scaling. The results of such a test is summarized in Fig.~\ref{fig5}.
The left panel of the figure shows a comparison of $v_2 (p_T)$ for protons, kaons and pions 
measured with the TOF detector for the centrality selection 5-30\%. The characteristic flavor 
dependence of $v_2$ is cleanly evidenced by these data, ie. mass ordering at low momentum 
and a reversal of the magnitudes of the $v_2$ values for baryons and mesons for  
$p_T \sim$ 1.8-2.0 GeV/c. The right panel of Fig.~\ref{fig5} and Fig.~\ref{fig6} show that 
very good scaling of $v_2$ is achieved with $y_T^{fs}=k_m\cdot  y_T^2 m$ in accordance 
with the predictions of hydrodynamics. Fig.~\ref{fig6} gives an illustration of y$_{T}^{fs}$ 
scaling for combined results which include the neutral kaons and lambda hyperons measured 
by the STAR collaboration. The latter were obtained in Au+Au collisions at 
\sqrtsNN~=200~GeV for the same centrality selection~\cite{star2}. 
Although the data shown in Fig.~\ref{fig6} indicate rather good  scaling for all measured 
particles over a broad range of y$_{T}^{fs}$, one can see clear evidence for a break in this
scaling for y$_{T}^{fs}\ge 2$. Such a break could be signaling a change in mechanism for 
high $p_T$ particles or a break down of ideal hydrodynamic flow. 


\subsection{Higher harmonic scaling}\label{harmonic_scaling}
\begin{figure}[htb!]
\hspace*{0.55cm}
\includegraphics[width=0.8\linewidth]{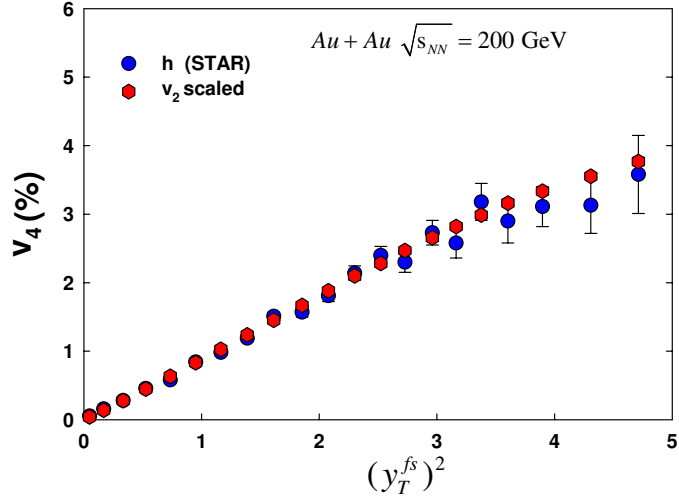}
\vskip -0.5cm
\caption[]{ $y_T^{fs}$ dependence for $v_4$ of charged particles from Au+Au collisions  at \sqrtsNN~=200~GeV
measured by STAR collaboration \cite{Adams:2003zg,Adams:2004bi} and these obtained by scaling $v_2$ values according to Eq.~\ref{eq5}
\label{fig7}} 
\end{figure}

As was pointed out earlier, ideal hydrodynamics \cite{buda-lund} gives  predicts 
a straightforward scaling relationship  between the second harmonic $v_2$ and the 
higher harmonics of azimuthal anisotropy ( $v_4$, $v_6$, ...). It is easy to show that  
the leading term of the relationship between $v_2$ and $v_4$ can can be expressed 
as
\begin{equation}
v_4=\frac{1}{2}\cdot v_2^2+ k_5 y_T^4,
\label{eq5} 
\end{equation}
 The measurement of the higher harmonics of azimuthal anisotropy are not available in PHENIX ( ongoing ).
However, we can test the predicted scaling relationship  between  $v_2$ and $v_4$ via data published 
by the STAR collaboration~\cite{Adams:2003zg,Adams:2004bi}. 
Figure \ref{fig7} shows the $(y_T^{fs})^2$ dependence of $v_4$ for charged particles 
from Au+Au collisions  at \sqrtsNN~=200~GeV and $v_2$ values scaled according to Eq.~\ref{eq5}.
The figure indicates rather good agreement  between the measured $v_4$ values and scaled $v_2$ 
values  over a broad range in $y_{T}^{fs}$. 

\section{Summary and Conclusions}\label{concl}
In summary detailed measurements of the fine structure of elliptic flow 
in Au+Au collisions at \sqrtsNN~=200 GeV measured by PHENIX 
collaboration at RHIC are presented. They show eccentricity scaling and $y_T^{fs}$ 
flavor scaling over a broad range of centralities and particle
flavors. This observed scaling gives strong support for essentially 
ideal hydrodynamical flow at RHIC. The observed scaling also supports the 
picture of a suddenly hadronizing (recombining) fluid of quarks.

% The scaling parameters can provide 
%important constraints for the thermodynamic and dynamical 
%properties of the high energy density matter created at RHIC. 

\section*{Acknowledgment(s)}
The author thanks R.A.  Lacey, M. Csan\'ad and  T. Cs\"org\H{o} for stimulating discussions.

\vfill\eject
\end{document}